\documentclass[12pt]{article}
\usepackage{amssymb,amsmath}
\usepackage{latexsym}
\usepackage{graphicx}
\usepackage{psfrag}

\numberwithin{equation}{section}

\def\coeff#1#2{\relax{\textstyle {#1 \over #2}}\displaystyle}

\def\IR{\mathbb{R}}

\def\cM{{\cal M}}
\def\cN{{\cal N}}

\def\cQ{{\cal Q}}

\newcommand{\be}{\begin{equation}}
\newcommand{\ee}{\end{equation}}
\newcommand{\bea}{\begin{eqnarray}}
\newcommand{\eea}{\end{eqnarray}}

\begin{document}

\begin{titlepage}

\begin{flushright}
IPhT-T09/130
\end{flushright}

\bigskip
\bigskip
\centerline{\Large \bf A (Running) Bolt for New Reasons}
\medskip
\bigskip
\centerline{{\bf Iosif Bena$^1$, Stefano Giusto$^1$,}}
\centerline{{\bf Cl\'{e}ment Ruef$^{\, 1}$ and Nicholas P. Warner$^2$}}
\bigskip
\centerline{$^1$ Institut de Physique Th\'eorique, }
\centerline{CEA Saclay, 91191 Gif sur Yvette, France}
\bigskip
\centerline{$^2$ Department of Physics and Astronomy}
\centerline{University of Southern California} \centerline{Los
Angeles, CA 90089, USA}
\bigskip
\centerline{{\rm iosif.bena@cea.fr,~stefano.giusto@cea.fr,~clement.ruef@cea.fr,~warner@usc
.edu }}

\bigskip \bigskip

\begin{abstract}

  We construct a four-parameter family of smooth, horizonless,
  stationary solutions of ungauged five-dimensional supergravity by
  using the four-dimensional Euclidean Schwarzschild metric as a base
  space and ``magnetizing'' its bolt.  We then generalize this to
  a five-parameter family based upon the Euclidean Kerr-Taub-Bolt.
  These ``running Bolt'' solutions are necessarily non-static.  They
  also have the same charges and mass as a non-extremal black hole
  with a classically-large horizon area. Moreover, in a certain regime
  their mass can decrease as their charges increase. The existence of
  these solutions supports the idea that the singularities of
  non-extremal black holes are resolved by low-mass modes that correct
  the singularity of the classical black hole solution on large
  (horizon-sized) scales.

\end{abstract}

\end{titlepage}


\section{Introduction}

Over the past few years there has been a significant shift in the
discussion of the ``fuzzball proposal'' and its realization in terms
of microstate geometries\footnote{Microstate geometries are defined to
  be smooth, horizonless solutions with the same asymptotic behavior
  at infinity as a black hole or black ring.}: Met initially with
considerable skepticism, there is now growing evidence (see
\cite{fuzzball-reviews} for reviews) that this proposal might well be
realized for BPS black holes. In hind-sight, this may not appear so
strange: extremal BPS black holes have a timelike singularity, and, as
is fairly well known, string theory oftentimes resolves such
singularities in terms of configurations that contain extra brane
dipole moments, and that have a size that is parametrically much
larger than the ``size'' of the original region of high
curvature\footnote{A few of the better-known examples of such
  singularity-resolution mechanism can be found in
  Polchinski-Strassler \cite{Polchinski:2000uf}, Klebanov-Strassler
  \cite{Klebanov:2000hb}, the D1-D5 system \cite{D1D5}, giant
  gravitons and LLM \cite{McGreevy:2000cw,LLM} or the massive M2
  supergravity dual \cite{m2,LLM}.}.

The BPS microstate geometries constructed thus far
\cite{microstate-papers} indicate that the timelike singularity of
extremal BPS black holes is resolved in a similar manner and that the
size of the configurations that resolve the singularity is of the same
order as the size of the black-hole horizon. This means that one can
no longer trust the ``classical'' space-time description of the region
between the timelike singularity and the horizon of the extremal black
hole, much as one does not trust the physical descriptions provided by
the Klebanov-Tseytlin solution \cite{kt}, the singular giant graviton,
or the unpolarized Polchinski-Strassler solution \cite{gppz} at scales
smaller than the singularity-resolution scale.  The resolution of
these singularities involves low-mass degrees of freedom, that affect
the physics at large distances; one cannot simply repair these
solutions in the neighborhood of the singularity by some Planck-scale,
or string-scale details.

The timelike singularity of BPS extremal black holes therefore appears
to be resolved in the same way as its string theory cousins that do
not sit behind a horizon, and this effectively implements the fuzzball
proposal for this class of black holes. If this proposal also applies
for {\it non-extremal} black holes, the would-be singularity
resolution mechanism would be even more remarkable: The singularity of
a non-extremal black hole is in the future of the horizon and if the
classical black hole is to be replaced by a superposition of
horizon-sized horizonless configurations this would imply that the
resolution of non-extremal black hole singularities will affect the
spacetime for a macroscopically-large distance in the past of the
singularity!  It is clearly important to understand this
singularity-resolution mechanism, not only because there is, as yet,
no rigorous example in string theory of how one might expect a
space-like singularity to be resolved, but also because we live in a
universe in which such singularities appear to be ubiquitous.

To establish that the singularity of non-extremal black holes is
resolved by horizon-sized horizonless geometries that have the same
mass and charges as the black hole, one first needs to construct such
geometries, which is no easy task -- only three such geometries are
known at present
\cite{Jejjala:2005yu,Giusto-Ross-Saxena,AlAlawi:2009qe} and some of their 
properties are studied in \cite{Gimon:2007ps}. One then
needs to see whether the physical properties of these geometries
support thinking about them as microstates of the non-extremal black
hole (and thus as examples of resolution of the black hole
singularity). For example, the geometry constructed in
\cite{Jejjala:2005yu}, which has an ergosphere but no horizon was
found to be unstable in \cite{Cardoso:2005gj} but the decay time was
then computed in the dual CFT \cite{samir-recent}, and found to match
exactly the decay time computed in gravity. This remarkable agreement
strongly supports thinking about the geometries of
\cite{Jejjala:2005yu} as microstates of non-extremal black holes, and
brings hope that the fuzzball proposal will equally apply to such
black holes.

Our purpose in this paper is to construct a new family of smooth,
horizonless solutions that have the same charges and mass as
non-extremal black holes, and that exist in the same regime of
parameters where the black hole exists. The solutions we find can be
thought of as asymptotically $\IR^{3,1} \times S^1$ solutions of
$U(1)^N$ supergravity in five dimensions, and from a four-dimensional
``string theory on Calabi-Yau'' perspective have D4, D2, D0 charges
and angular momentum.  Smoothness requires some of these charges to be
related, and the solution depends on $N+2$ independent parameters.

Unlike the non-extremal solutions of
\cite{Jejjala:2005yu,Giusto-Ross-Saxena,AlAlawi:2009qe}, these
solutions do have the same charges and mass as a black hole with a
classically-large horizon area. Furthermore, they are asymptotically
flat, and do not contain an $AdS$ region.  This is both an advantage
-- they can be microstates of more generic non-extremal black holes,
and a disadvantage -- one cannot use the power of the $AdS$-CFT
correspondence to study them.  Consequently, the interpretation of
these geometries as microstate geometries of a black hole will be
somewhat less compelling than it has been for the BPS microstates.

The technical construction of these non-extremal geometries is not as
complicated as one might expect, and is in fact very similar to the
construction of BPS and non-BPS extremal multi-center solutions. To
construct the latter one takes a four-dimensional hyper-K\"ahler
manifold with self-dual curvature, turns on self-dual
\cite{5dsugra,gutowski-reall,Bena:2004de} two-forms (for BPS
solutions) or anti-self dual \cite{Goldstein:2008fq} two-forms (for
non-BPS solutions), and solves a linear system of equations
\cite{Bena:2004de} to determine the warp factors and the angular
momentum. The nice observation of \cite{Goldstein:2008fq} that one can
find ``almost-BPS'' non-supersymmetric solutions by simply flipping
some relative orientations has been further exploited in
\cite{Bena:2009ev,recent} where large classes of such solutions have
been constructed. One way to think about these solutions is as coming
from supersymmetric brane configurations that live in gravitational
backgrounds that are incompatible with the supersymmetries of the
branes.

One can, however, take this idea even further and ask whether the
hyper-K\"ahler condition is really necessary if one only wishes to
satisfy the supergravity equations of motion.  Of course, having a
hyper-K\"ahler base was needed for supersymmetry, but Einstein's
equations should only care about the Ricci tensor of the base and
whether the base is Ricci-flat or not.  Hence, if we consider the BPS
equations of \cite{5dsugra,gutowski-reall,Bena:2004de}, or the non-BPS
equations of \cite{Goldstein:2008fq}, and replace the hyper-K\"ahler
base by any Ricci-flat base we still expect to obtain a (non-BPS)
solution to the equations of motion.  In \cite{Eoms-IW} we prove this
result in detail, starting from the full equations of motion for
five-dimensional supergravity coupled to three U(1) gauge fields.

Our purpose in this paper is to study some examples of this idea, and
to find smooth, horizonless solutions in five dimensions using
Ricci-flat, Euclidean four-dimensional base metrics.  Perhaps the
simplest, most interesting such metrics arise from the
Euclideanization of the various pure-gravity black-hole metrics.  Such
solutions have a periodic ``imaginary time'' coordinate and are thus
asymptotic to $\IR^3 \times S^1$. These solutions also come with a
``bolt,'' that is, the center of these solutions is topologically
$\IR^2 \times S^2$ where the $S^2$ remains of finite size.  This is
because the Euclideanized geometry closes off smoothly where the
(outer) horizon of the original Lorentzian black hole used to be and
the $S^2$ at the center is the same size as the original black-hole
horizon.  We therefore use these Euclidean metrics as base metrics for
five-dimensional solutions and then add smooth (cohomological) self-dual or
anti-self-dual magnetic fluxes to the bolt, and these fluxes also act as
sources for the electric fields and angular momentum.

While the manner of obtaining these solutions is similar to that of
BPS solutions, there are some fundamental differences.  First, and
most obviously, these solutions are not BPS.  In addition, their mass
depends on a combination of the electric charges and the mass
parameter of the underlying Euclidean black hole. Furthermore, the
smooth BPS bubbled geometries have an ambi-polar\footnote{This means
  that the auxiliary, four-dimensional base metric can change
  signature from $+4$ to $-4$.  The physical five-dimensional metric
  constructed from such a base will, however, be smooth and Lorentzian
  \cite{Giusto:2004kj,Bena:2005va,Berglund:2005vb,Saxena:2005uk}.}
base-space that by itself is rather pathological, and if one sets to
zero some of the fluxes, the two-cycles of the base collapse. In
contrast, the non-BPS solutions considered here have base spaces that
can be either regular or ambi-polar, and that can give regular
five-dimensional solutions even in the absence of fluxes. The
two-cycle of the base space is the bolt.  ``Magnetizing'' the bolt, by
adding fluxes, distorts the size of this bolt and of the $S^1$ at
infinity but removing the flux does not collapse the bolt.
Furthermore, when fluxes are added, the Euclidean time direction of
the base (which is now the Kaluza-Klein direction) and the physical
time direction mix. Hence these solutions are never static at infinity
in the frame where the metric is static at the bolt, and viceversa.
We therefore refer to them as ``running Bolt''
solutions\footnote{Without, of course, any reference to the recent
  breaking of the 100 and 200 meters world records.}.

Before beginning we should also note that most, if not all the
five-dimensional solutions that result from taking an Euclidean
instanton and simply adding time\footnote{Solutions of this type are usually referred to in the literature as
static KK bubbles. For a general discussion of this class of solutions see, for example, \cite{KKbubbles}.} are unstable. This was demonstrated
for the Euclidean Schwarzschild solution by Gross, Perry and Yaffe
\cite{GPY}, and for the Taub-bolt instanton by Young \cite{Young:1983dn}. The
more general Kerr-Taub-bolt instanton has not yet, to our
knowledge,  been found to have negative modes. The solutions we construct
also have magnetic and electric fluxes, as well as angular momentum,
and their stability analysis is likely to be more complicated than
that of \cite{GPY,Young:1983dn} (which is not simple either). By continuity, it is quite
likely that the running Bolt solutions will still be unstable when the fluxes are
small but when the fluxes and charges become larger the fate of the solutions is
unknown -- they may become stable or remain unstable. However, this
need not hamper their interpretation as microstates of the
corresponding black hole.  On the contrary, as for the microstates of
\cite{Jejjala:2005yu}, the instability may be necessary from a
microscopic perspective \cite{samir-recent}, and the variation of the
decay time with the charges of our solutions might as well give a way
to identify the black hole microstates that correspond to our
solutions.

In Section 2 we give our conventions and specify the class of
solutions we are going to consider.  In Section 3 we generate new
solutions from the Euclidean Schwarzschild metric.  This solution is
extremely simple since spherical symmetry is preserved throughout.  We
then go on, in Section 4, to consider solutions generated from the the
Euclidean Kerr-Taub-Bolt metric.  While these solutions have some
similar general features to the solution based upon the Schwarzschild
metric, the Kerr-Taub-Bolt solution is richer and has more parameters.
Indeed, {\it a priori}, this solution has three independent parameters
but the combined effect of removing   conical singularities and Dirac strings
 in the Euclidean base imposes a cubic constraint on
the parameters and this implies that only two of them are independent.  There are, however, potentially several
branches in the solution space, but physical conditions, like the requirement of a
positive definite Kerr-Taub-Bolt metric mean that one must discard
some of the branches and parameter ranges \cite{Gibbons:1979nf}.  This
raises the interesting question of whether, by allowing an ambi-polar
base metric, one can generate more solutions than those allowed by the naive positive-definite
constraint on the base metric.  We find that this is indeed possible, and that not only is there a broader range of
admissible parameters when one allows ambi-polar base metrics, but that there are additional branches to the
smooth five-dimensional solutions coming from allowable changes of sign in the cubic
constraint.
Finally, we conclude Section 4 with a computation of the
asymptotic charges of running Bolt solutions, and Section 5 contains some further remarks.

\section{The family of solutions}

\subsection{Conventions}

We consider $\cN \! = \!  2$, five-dimensional supergravity with three $U(1)$ gauge fields and we use the conventions of \cite{Goldstein:2008fq}.  The bosonic action may be written as:
\begin{eqnarray}
  S = \frac {1}{ 2 \kappa_{5}} \int\!\sqrt{-g}\,d^5x \Big( R  -\coeff{1}{2} Q_{IJ} F_{\mu \nu}^I   F^{J \mu \nu} - Q_{IJ} \partial_\mu X^I  \partial^\mu X^J -\coeff {1}{24} C_{IJK} F^I_{ \mu \nu} F^J_{\rho\sigma} A^K_{\lambda} \bar\epsilon^{\mu\nu\rho\sigma\lambda}\Big) \,,
  \label{5daction}
\end{eqnarray}
with $I, J =1,2,3$.   One of the photons lies in the gravity multiplet and so there are only two vector multiplets and hence only two independent scalars.  Thus the scalars, $X^I$, satisfy a constraint, and it is convenient to introduce three other scalar fields, $Z_I$, to parameterize these two scalars:
\begin{equation}
X^1 X^2 X^3  = 1\,, \qquad  X^1    =\bigg( \frac{Z_2 \, Z_3}{Z_1^2} \bigg)^{1/3} \,, \quad X^2    = \bigg( \frac{Z_1 \, Z_3}{Z_2^2} \bigg)^{1/3} \,,\quad X^3   =\bigg( \frac{Z_1 \, Z_2}{Z_3^2} \bigg)^{1/3}  \,.
\label{XZrelns}
\end{equation}
The metric for the kinetic terms can be written as:
\begin{equation}
  Q_{IJ} ~=~    \frac{1}{2} \,{\rm diag}\,\big((X^1)^{-2} , (X^2)^{-2},(X^3)^{-2} \big) \,.
\label{scalarkinterm}
\end{equation}
Note that the scalars, $X^I$,   only depend upon the ratios $Z_J/Z_K$ and it is convenient to parameterize a  third independent scalar by:
\begin{equation}
Z ~\equiv~ \big( Z_1 \, Z_2 \, Z_3  \big)^{1/3}   \,.
\label{Zdefn}
\end{equation}

We now use this scalar, $Z$, in the metric Ansatz:
\begin{equation}
ds_5^2 ~=~ -Z^{-2} \,(dt + k)^2 ~+~ Z \, ds_4^2  \,,
\label{metAnsatz}
\end{equation}
where the powers guarantee that $Z$ becomes an independent scalar from the four-dimensional perspective.
We will denote the frames for (\ref{metAnsatz}) by $e^A$, $A=0,1, \dots,4$  and let $\hat e^a$, $a=1, \dots,4$ denote frames for $ds_4^2$.  That is, we take:
\begin{equation}
e^0 ~\equiv~     Z^{-1} \,(dt + k)\,, \qquad e^a ~\equiv~  Z^{1/2} \,\hat e^a \,.
\label{frames}
\end{equation}

The Maxwell Ansatz is:
\begin{equation}
A^I   ~=~  - \varepsilon\, Z_I^{-1}\, (dt +k) + B^{(I)}  \,,
\label{AAnsatz}
\end{equation}
where $B^{(I)}$ is a one-form on the base (with metric $ds_4^2$). The parameter, $\varepsilon$, will be related to the self-duality or anti-self-duality of the fields in the solution  and is fixed  to have $\varepsilon^2 =1$.  It is convenient to define the field strengths:
\begin{equation}
\Theta^{(I)}    ~\equiv~  d B^{(I)}  ~=~ \coeff{1}{2} \, Z^{-1} \,\Theta^{(I)}_{ab} \, e^a \wedge e^b    ~=~ \coeff{1}{2} \, \Theta^{(I)}_{ab} \, \hat e^a \wedge \hat e^b   \,.
\label{Thetadefn}
\end{equation}
Note that the frame components are defined relative to the frames on $ds_4^2$.

\subsection{BPS and simple non-BPS solutions}

Both the BPS and the almost-BPS solutions are given by taking the base to be hyper-K\"ahler with a self-dual curvature and then solving the linear system \cite{Bena:2004de,Goldstein:2008fq}:
\begin{eqnarray}
\Theta^{(I)} &=&  \varepsilon *_4 \Theta^{(I)} \,,  \label{BPSeqna} \\  \hat \nabla^2 Z_I &=&   \coeff{1}{2} \, \varepsilon \, C_{IJK}   *_4 [\Theta^{(J)} \wedge \Theta^{(K)}] \,,\label{BPSeqnb} \\
 d k ~+~\varepsilon  *_4 d k  &=&   \varepsilon \,Z_I \Theta_I  \label{BPSeqnc}     \,.
\end{eqnarray}
As we explain in \cite{Eoms-IW}, one can obtain solutions to the equations of motion simply by solving the BPS system (\ref{BPSeqna})--(\ref{BPSeqnb}) with any Ricci-flat base metric on $ds_4^2$:
\begin{equation}
 \hat R_{ab}  ~=~ 0   \,.
\label{Ricciflat}
\end{equation}
 The most obvious such base is the Euclidean Schwarzschild metric, which we use in the next section to generate new solutions. In Section 4 we will extend this to the more general Kerr-Taub-bolt solution.

\section{Adding fluxes to Euclidean Schwarzschild}

\subsection{The solution}

The Euclidean Schwarzschild metric is given by:
\begin{equation}
ds_4^2 ~=~ \Big(1- {2m \over r}\Big) \, d \tau^2 ~+~ \Big(1- {2m \over r}\Big)^{-1}\, d r^2~+~ r^2 d \theta^2 ~+~ r^2 \sin^2 \theta \, d \phi^2 \,.
\label{EuclSch}
\end{equation}
It is, of course, Ricci flat, and if one restricts to the region $r \ge 2 m$ then the metric is globally regular provided one periodically identifies the Euclidean time  by:
\begin{equation}
\tau ~\equiv~ \tau ~+~ 8 \pi  m  \,.
\label{SchPer}
\end{equation}
Near $r=2m$ the manifold is then locally $\IR^2 \times S^2$ and at infinity it is asymptotic to $\IR^3 \times S^1$.
The ``bolt'' at the origin can be given a magnetic flux  and we can take the $\varepsilon$-self-dual harmonic two-forms to be:
\begin{equation}
\Theta^{(I)} ~=~ q_I \, \Big(  {1 \over r^2} \, d \tau  \wedge dr ~+~  \varepsilon\, \sin  \theta \, d \theta  \wedge d \phi  \Big)   \,,
\label{harmtwoform}
\end{equation}
for some magnetic charges, $q_I$.

With this flux it is trivial to solve the second equation (\ref{BPSeqnb}) and one finds
\begin{equation}
Z_I ~=~  1 ~-~ \coeff{1}{2} \, C_{IJK}  \, {q_J q_K \over  m} \, {1 \over  r}  \,.
\label{ZISch}
\end{equation}
We have chosen the homogeneous solution so as to exclude all other electric sources for $Z_I$ and to arrange that
$Z_I \to 1$ as $r \to \infty$.

The last equation   (\ref{BPSeqnc}) is equally elementary, and setting
\begin{equation}
k ~=~ \mu   \, d\tau  ~+~ \omega \, d \phi  \,
\label{kSch}
\end{equation}
we find
\begin{eqnarray}
\mu  &=&   (\varepsilon + \alpha) \, (q_1 + q_2 + q_3) \, {1 \over r}  ~-~ {3\varepsilon \over 2 m} \,  q_1 q_2 q_3  \, {1 \over  r^2} ~+~ \gamma   \,, \label{Schmu} \\
  \omega &=& \alpha (q_1 + q_2 + q_3) (\beta + \cos \theta)  \,,
\label{Schom}
\end{eqnarray}
where $\alpha, \beta$ and $\gamma$ parameterize homogeneous solutions.  These parameters must be chosen to  remove closed time-like curves (CTC's) in  (\ref{metAnsatz}).  First, to avoid CTC's on $\phi$-circles one must make sure that there are no Dirac strings in $k$, and hence $\alpha = \beta =0$.  Similarly, there are potential CTC's around the small $\tau$ circles near $r = 2m$ unless we choose $\gamma$ so that $\mu = 0$ at $r=2m$. Thus we must take $ \omega = 0$ and
\begin{equation}
\mu  ~=~     \varepsilon (q_1 + q_2 + q_3) \, \Big({1 \over r} -{1 \over 2m} \Big)  ~-~ {3\varepsilon\over 2 m}  \,    q_1 q_2 q_3  \,  \Big( {1 \over  r^2} -{1 \over 4m^2} \Big)   \,.  \label{Schmures} \\
\end{equation}

The solution is now completely determined but it still remains to verify the absence of CTC's elsewhere. On the constant time slices the  metric in the $\tau$ direction is $\cM d \tau^2$ where $\cM   \equiv  Z^{-2}  r^{-4}   \cQ $ and
\begin{equation}
\cQ ~\equiv~ r^4   Z_1 Z_2 Z_3  \Big(1- {2m\over r} \Big) ~-~ \mu^2 r^4  \,.  \label{CTCtest} \\
\end{equation}
This is a quartic function of $r$ and must remain non-negative for $2m \le  r < \infty$ and this places constraints on $m$ and the $q_I$.   In addition, the $Z_I$ should remain positive definite for $r > 2m$.

To examine these conditions in more detail we simplify the analysis by taking $q_I = q > 0$, $I=1,2,3$.  The positivity of the $Z_I$ for $r > 2m$ means that one must have:
\begin{equation}
q ~< ~\sqrt{2}\, m   \,. \label{simpchcond} \\
\end{equation}
To analyze the positivity of $\cM$, we first look at the behavior at infinity, where one has
\begin{equation}
\cM ~\sim~ r^{-4} \cQ ~ \sim ~     \Big( 1  -  {3 q \over 8 m^3 }   (q^2 - 4  m^2 ) \Big) \Big( 1  +   {3 q \over 8 m^3 }   (q^2 - 4  m^2 ) \Big) \,.  \label{CTCinf} \\
\end{equation}
For this to be positive, the two cubics in $q$ must be positive and this implies the stronger condition:
\begin{equation}
  0 ~<~  {q\over m}   ~<~ {4 \over \sqrt{3}} \, \sin {  \pi \over 9} ~\approx~ 0.78986 \,.  \label{chcond1} 
\end{equation}
Note that the function $\cM$, and hence the condition above, do not depend on $\varepsilon$.
More generally, the quartic that sets the scale of the $\tau$-circle is:
\begin{equation}
\cQ  ~\equiv~   (r - 2m) \Big[  \Big(  r  - {q^2 \over m} \Big)^3  ~-~    {9 q^2 \over 64 m^6 }   (r - 2m)  \big(  (q^2 - 4 m^2 ) r  + 2 m q^2 \big)^2  \Big] \,.  \label{CTCtestsimp} \\
\end{equation}
One can verify that this is indeed positive definite for $2m < r < \infty$ for $q$ in the range (\ref{chcond1}).

\begin{figure}[t]
 \centering
    \includegraphics[width=10cm]{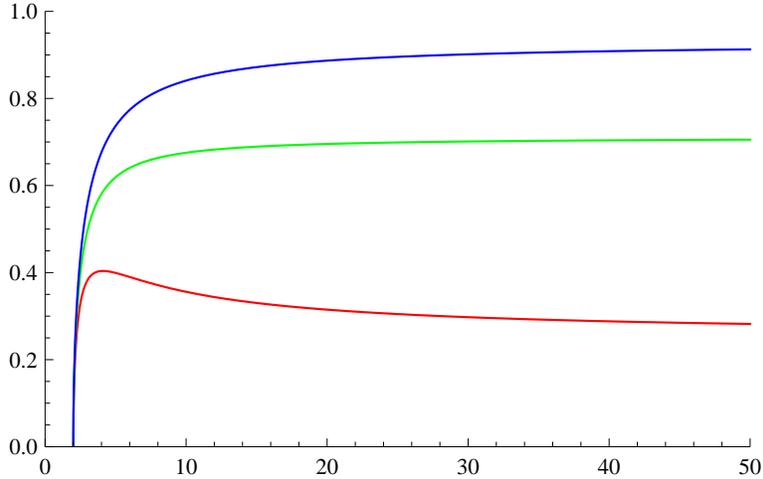}
    \caption{\it \small  Plot of the scale, $\sqrt{\cM }$,  of the compactification circle as a function of $r/m$. The three plots, from top to bottom,  correspond to $q/m$ of $1/4$, $1/2$ and $3/4$.  Note that as one approaches the upper bound (\ref{chcond1}) the circle does not grow uniformly but attains a maximum scale before decreasing asymptotically.}
\label{fig1}
\end{figure}

It is interesting to note that (\ref{Schmures}) shows that $\mu$
asymptotes to a finite value as $r \to \infty$.  One can undo the
rotation of this frame by shifting $\tau \to \tau + a t$ and the
condition (\ref{chcond1}) simply reflects the fact
that this rotation is sub-luminal.

We have thus created a ``magnetized bolt'' solution in which fluxes
have been added to a pre-existing two-cycle. It is interesting to note
that in the BPS ``bubbled'' solutions of
\cite{Bena:2005va,Berglund:2005vb,Saxena:2005uk,Balasubramanian:2006gi} the fluxes were an
essential part of blowing up the two-cycles and these bubbles would
collapse without the fluxes.  Another element of the BPS bubbled
solutions was the presence of an ambi-polar base metric where the
metric on the four-dimensional base changes sign but this sign change
is canceled in the five-dimensional metric by a simultaneous sign
change in warp factor, $Z$.  In more physical terms, there is also a
direct D-brane interpretation of the bubbling transition
\cite{Bena:2005va,Balasubramanian:2006gi}.

The solution constructed here does not appear to have such a D-brane
interpretation and does not involve an ambi-polar base.  One could try
to see whether the ranges of parameters or the range of $r$ might be
extended to give an ambi-polar four-dimensional base that still yields
a smooth Lorentzian five-dimensional solution.  There are obvious
possibilities, like taking $m<0$ and trying to extend to $r<0$ but
such extensions do not lead to an {\it overall} sign change in
(\ref{EuclSch}) and so cannot be canceled by the warp-factor $Z$.
Thus, at least for this solution, the standard Euclidean Bolt is
simply decorated by fluxes to give a running Bolt solution with
electric and magnetic charges.  We shall see later that there are
richer possibilities once angular momentum and a NUT charge are
included.

\subsection{Asymptotic Charges}
\label{sec:charge}
The solution carries M5 charges, which are encoded in the magnetic part of the gauge field, $B^{(I)}$, and are equal to $q_I$. The solution also carries M2 charges. Note that   the gauge field equations involve Chern-Simons terms:
\be
d  ((X^I)^{-2} *_5 d A^I)= \coeff{1}{2}  C_{IJK} dA^J \wedge dA^K\,.
\ee
In the presence of such terms, the proper definition of the conserved electric charge associated with $A^I$ is
\be
Q^I =  \int_{S^1\times S^2}  \Bigl[(X^I)^{-2} *_5 d A^I-\coeff{1}{2} C_{IJK} A^J \wedge d A^K\Bigr]\,,
\label{QI}
\ee
where the integral is computed over the $S^1$ circle parameterized by $\tau$ and the $S^2$ sphere at spatial infinity. The Chern-Simons term gives a non-vanishing contribution to the charge, due to the fact that the one-form $k$ goes to a constant non-zero value at infinity:
\be
k\to \gamma d\tau\,,\quad \gamma =   -\varepsilon {q_1 + q_2 + q_3\over 2 m}+3\varepsilon \,{q_1 q_2 q_3\over 8 m^3}\,.
\ee
Using the identity
\bea
&&(X^I)^{-2} *_5 d A^I  -\coeff{1}{2}  C_{IJK} A^J \wedge d A^K  \\ && \qquad \qquad \qquad~=~
\varepsilon *_4 d Z_I -\coeff{1}{2}  C_{IJK} B^{(J)} \wedge \Theta^{(K)}
+{\varepsilon\over 2} C_{IJK} d \Bigl[(dt+k)\wedge {B^{(J)}\over Z_K}\Bigr]\, ,\nonumber
\eea
one finds
\be
 Q^I =- (8\pi m) (4\pi)  \coeff{1}{2} \, C_{IJK} \Bigl[ \varepsilon {q_J q_K \over  m} + {\gamma\over 2} (q_J+q_K)\Bigr]\,.
\label{charges}
\ee

To compute the mass and the KK electric charge of the solution one has
to analyze the asymptotic form of the metric. The fact that the
one-form, $k$, does not vanish at infinity implies that the
coordinates $(\tau,t)$ define a frame which is not asymptotically at
rest, much like for the Black Ring in Taub-NUT constructed in
\cite{Elvang:2005sa}. One can go to an asymptotically static frame by
re-writing the large $r$ limit of the metric in the form
\be
ds^2\approx (1-\gamma^2)\Bigl(d\tau - {\gamma\over 1-\gamma^2} dt \Bigr)^2-dt^2 (1-\gamma^2)^{-1} + dr^2 + r^2 (d\theta^2+\sin^2\theta d\phi^2)\,,
\ee
and redefining the coordinates as
\be
\hat{\tau}=(1-\gamma^2)^{1/2} \Bigl(\tau -{\gamma\over 1-\gamma^2} t\Bigr)\,,\quad \hat{t}=(1-\gamma^2)^{-1/2} t\,.
\label{hatcoord}
\ee
The condition (\ref{chcond1}) reflects the fact
that the rotation is sub-luminal ($\gamma<1$) and hence this change of
coordinates is well-defined.

Dimensional reduction of the five-dimensional metric along the direction $\hat \tau$ yields
\bea
ds^2_5=\Bigl(1-{2 m\over r}\Bigr)^2 Z^{-2} \hat{I}_4\Bigl[d\hat \tau - \Bigl(1-{2 m\over r}\Bigr)^{-2}\mu \hat{I}^{-1}_4 d\hat t + \gamma d\hat t\Bigr]^2+ \Bigl(1-{2 m\over r}\Bigr)^{-1} Z \hat{I}_4^{-1/2} ds^2_E\,,
\eea
where
\be
\hat{I}_4= (1-\gamma^2)^{-1}\Bigl[Z^3 \Bigl(1-{2 m\over r}\Bigr)^{-1} - \mu^2  \Bigl(1-{2 m\over r}\Bigr)^{-2}\Bigr] \,,
\ee
and
\be
ds^2_E= -\hat{I}_4^{-1/2} d\hat{t}^2 + \hat{I}_4^{1/2} \Bigr[dr^2 + \Bigl(1-{2 m\over r}\Bigr) (r^2 d\theta^2 + r^2 \sin^2\theta d\phi^2)\Bigr]
\ee
is the four-dimensional Einstein metric. From the coefficient of $d\hat{t}^2$ in the Einstein metric one can read off the mass
of the solution:
\be
G_4 M = {1\over(1- \gamma^2)}\Bigl({m\over 2} - {q_1q_2+q_1q_3+q_2q_3 \over 4m} - {3\over 8}{\gamma\over \varepsilon} {q_1 q_2 q_3\over m^2}\Bigr)\,.
\ee
Here $G_4$ is the four-dimensional Newton's constant, whose relation with the five-dimensional Newton's constant $G_5$ is
\be
G_4 = {G_5\over (1-\gamma^2)^{1/2} (8\pi m)}\,.
\ee
The KK electric charge $Q_e$ is encoded in the KK gauge field
\be
A_{KK}=  \left(\gamma - \Bigl(1-{2 m\over r}\Bigr)^{-2}\mu \hat{I}^{-1}_4 \right) d\hat{t}
\ee
and is given by
\be
G_4 Q_e=-{1\over 4(1-\gamma^2)}\Bigl({3\varepsilon\over 4} {q_1 q_2 q_3\over m^2}+\gamma {q_1 q_2 +q_1 q_3 +q_2 q_3\over m}+\varepsilon \gamma^2 (q_1 +q_2 +q_3)\Bigr)\,.
\ee

If one now computes the rest-mass, $M_0$, of the solution (i.e. the mass with respect to the ($t,\tau$) frame) one obtains:
\be
	M_0 \equiv (1-\gamma^2)^{-1/2}(M-\gamma Q_e) = {\pi \over 4 G_5}\left(16m^2 +{\varepsilon \over  4\pi^2} (Q^1+Q^2+Q^3 ) \right)\,.
\label{mass}
\ee

\subsection{Some Remarks on the Mass of the Running Bolt}
\label{remarks}

For $\varepsilon=1$, Equation (\ref{mass}) indicates that the total
rest mass is simply the sum of the mass of the uncharged bolt and the
masses corresponding to the M2 branes. Hence, if one could ascribe a
putative solitonic charge to the uncharged bolt, this formula would
look very much like the mass of a BPS object. Furthermore, the fact
that the M2 brane charge enters linearly in the total mass is also
consistent with the fact that a probe M2 brane feels no force in this
background.

For $\varepsilon=-1$ the situation is even more interesting. The mass
now decreases linearly with increasing the M2 charge. Hence, the mass
formula is still linear, but the sign in front of the M2 charges is
negative!  We are not aware of any other such mass formula in the
literature.  One might object to this by noting that one can always
flip the sign of M2 charges by reversing some orientations; however, by flipping the signs of some of the $q_I$ one can change the sign of some of the M2 charges.
Hence the total mass can either decrease or increase with
increasing the mass corresponding to the M2 charges. Alternatively, if one absorbes $\varepsilon$ by an orientation change,
then the mass formula (3.29) is linear in the charges, and not in their absolute values as for BPS systems.

The fact that the mass of the solutions can decrease with increasing charge
and the fact that M2 brane probes feel no force may lead one to believe
naively that one could violate energy conservation: one can bring an M2 brane adiabatically
from the infinity to the core of a solution, and the resulting solution will have a
lower mass than the sum of the masses of the two pre-merger components. However, this does
not happen. The charge of the M2 brane probe that feels no force is oriented oppositely to the
M2 brane charge of this solution! Bringing in this probe
brane actually decreases the total M2 charge and therefore increases the mass of the solution to
the mass of the soliton plus the mass of the probe M2 one brought in adiabatically, as expected.

Our analysis thus indicates that the uncharged bolt is the middle
point of a family of magnetized solutions that can have both larger
and smaller rest masses, and that moreover these masses can grow or
decrease linearly with the M2 charges of the solution.

\section{Adding fluxes to the Kerr-Taub-Bolt solution}

The Euclidean Schwarzschild solution admits a very interesting
generalization with additional quantum numbers: The Euclidean
Kerr-Taub-Bolt solution \cite{Gibbons:1979nf}. In this section we add
fluxes to the bolt and get a more general regular five-dimensional
running Bolt solution.

\subsection{The  Euclidean Kerr-Taub-Bolt Solution}

The four-dimensional metric is
\begin{equation}
ds_4^2 ~=~ \Xi \Big( {dr^2 \over \Delta}  ~+~ d \theta^2 \Big) \,  ~+~ {\sin^2  \theta \over \Xi}  \, (\alpha d \tau + P_r d \phi)^2  ~+~ {\Delta \over \Xi}  \, (d \tau + P_\theta d \phi)^2 \,,
\label{KTBmet}
\end{equation}
where
\begin{eqnarray}
\Delta  & \equiv&    r^2 - 2 m r - \alpha^2 + N^2   \,, \qquad \qquad \Xi ~\equiv~  P_r - \alpha P_\theta ~=~ r^2 - (N + \alpha \cos \theta)^2 \,,  \nonumber \\
 P_r & \equiv&   r^2 - \alpha^2 - {N^4 \over N^2 - \alpha^2}  \,, \qquad P_\theta   ~\equiv~  -\alpha \sin^2 \theta + 2 N \cos  \theta  -     {\alpha  N^2 \over N^2 - \alpha^2}  \,.
\label{KTBdefns}
\end{eqnarray}
This is a Ricci-flat metric where $m$ is the mass, $\alpha$ is the angular momentum and $N$ is the NUT charge. If the metric is to be regular then these parameters are not independent, as we will see in the following. At infinity the metric  behaves as
\begin{equation}
ds_4^2 ~\sim~ dr^2~+~  r^2 ( d \theta^2  + \sin^2  \theta \,  d \phi^2) ~+~ \big(d \tilde \tau  -( \alpha \sin^2 \theta + 2N (1-\cos \theta) )\, d \phi \big)^2 \,,
\label{asympKTBmet}
\end{equation}
with
\begin{equation}
  \tilde \tau ~\equiv~  \tau ~+~ 2N \phi ~-~  {\alpha  N^2 \over N^2 - \alpha^2} \, \phi   \,.
\label{tildetaudefn}
\end{equation}
Thus the metric is asymptotic to $\IR^3 \times S^1$ provided that $\phi$ has period $2 \pi$ and the fibration of $\tau$ over the two-sphere in $\IR^3$ is regular if $\tau$ is identified under shifts:
\begin{equation}
 \tau ~\equiv~  \tau ~+~ 8 N \pi    \,.
\label{tauident}
\end{equation}

We now need to examine regularity at the points where some of the metric coefficients vanish. First, at $\theta = 0,\pi$, the circle with $d \tau = - P_\theta d \phi$ pinches off. Substituting $d \tau = - P_\theta d \phi$ into the metric and ignoring the radial terms gives a metric:
\begin{equation}
\Xi  \,\Big[  d\theta ^2 ~+~ {1 \over \Xi^2}  \sin^2 \theta \, \big( (P_r - \alpha P_\theta) \, d \phi \big)^2 \Big]  ~=~
\Xi  \,\big[  d\theta ^2 ~+~   \sin^2 \theta \, d \phi ^2 \big]  \,,
\label{spheremet}
\end{equation}
which is perfectly regular.

The second degeneracy appears at $\Delta =0$. Define $r_\pm$ by  $\Delta = (r- r_+)(r-r_-)$ with $r_+ > r_-$:
\begin{equation}
r_\pm ~=~  m ~\pm~ \sqrt{m^2 - N^2 + \alpha^2}   \,.
\label{rpmdefn}
\end{equation}
Since we are interested in  Euclidean black-hole solutions with non-trivial bolts, we will consider the situation where these roots are real:
\be \label{rplusreal}
m^2 + \alpha^2 \geq N^2\,.
\ee
We have to arrange that the metric is regular at $r\rightarrow r_+$
and then restrict to $r \ge r_+$.  As usual, this will lead to a
periodic identification in the $\tau$ coordinate.  Before proceeding
with the analysis here, it is worth noting that in the usual analysis
of the Kerr-Taub-Bolt metric \cite{Gibbons:1979nf}, one requires the
metric to be positive definite and hence one requires $\Xi > 0$ and
hence $m>|N|$.

However, if one's purpose is to construct five-dimensional solutions,
the four-dimensional base can be ambi-polar and hence $\Xi$ {\it can}
be allowed to change sign. We will indeed find that the warp factors
also change sign to compensate for this and give a regular
five-dimensional solution. Thus we will not impose that $m>|N|$, but
only (\ref{rplusreal}).

To explore regularity at $r= r_+$ it is useful to define:
\begin{equation}
P_{r+}  ~\equiv~ P_r |_{r = r_+}  ~=~   r_+^2 - \alpha^2 - {N^4 \over N^2 - \alpha^2} \,, \qquad   \kappa ~\equiv~  \Big|{r_+ - r_- \over 2 P_{r+}}\Big|  \,.
\label{horvals}
\end{equation}
The circle that pinches off at $r= r_+$ has $d \phi = - \alpha d\tau/P_{r+}$.  Substituting this into the metric and expanding in  $x =(r-r_+)^{1/2}$ one obtains:
\begin{equation}
{4\, \Xi \over r_+ - r_-}  \,\Big[  dx^2 ~+~ \kappa^2 x^2 d \tau^2 ~+~ \coeff{1}{4} (r_+ - r_-) \, d \theta^2  \Big]   \,.
\label{nearhor}
\end{equation}
This is regular as $x \to 0$ provided that $\tau$ is periodically identified according to:
\begin{equation}
 \tau ~\equiv~  \tau ~+~ {2 \pi \over \kappa}    \,,
\label{othertauident}
\end{equation}
and hence the base space is smooth if
\begin{equation}
\kappa ~=~    {1  \over 4 | N |}    \,.
\label{paramreln}
\end{equation}
This condition, together with (\ref{rplusreal}) are the two necessary
condition for absence of conical singularities in the base. Hence,
the ambi-polar Kerr-Taub-Bolt metrics that we will
use to generate running Bolt solutions depend on only two independent parameters.

We now
explore the implications of equation (\ref{paramreln}) for the allowed
range of $N,m$ and $\alpha$. Since this equation only involves $|N|$,
one can use the definition of $\kappa$ to see that the sign of $N$ is
irrelevant; hence we will assume in the following that $N$ is positive.

One can now square the square roots in (\ref{paramreln}), and
obtain a constraint that is cubic in $m$. This constraint
depends on the sign of $P_{r+}$: if $P_{r+}$ is positive, we have
\begin{eqnarray}
&& 16 N(N^2 - \alpha^2)^2 m^3 - 4 (5N^6 -8 \alpha^2 N^4 + 2 \alpha^4 N^2 + \alpha^6) m^2  \nonumber \\
&& \quad - 16 N(N^2 - \alpha^2)^3 m + 20 N^8 - 52 \alpha^2 N^6  + 49 \alpha^4 N^4 - 16 \alpha^6 N^2 ~=~ 0 \,,
\label{cubic1}
\end{eqnarray}
while if $P_{r+}$ is negative  equation (\ref{paramreln}) implies:
\begin{eqnarray}
&& -16 N(N^2 - \alpha^2)^2 m^3 - 4 (5N^6 -8 \alpha^2 N^4 + 2 \alpha^4 N^2 + \alpha^6) m^2  
\nonumber \\
&& \quad + 16 N(N^2 - \alpha^2)^3 m + 20 N^8 - 52 \alpha^2 N^6  + 49 \alpha^4 N^4 - 16 \alpha^6 N^2 ~=~ 0 \,,
\label{cubic2}
\end{eqnarray}
which is the same as (\ref{cubic1}) but with $ m \to -m$.  Note that a
solution to (\ref{cubic1}) or (\ref{cubic2}) is not automatically a
solution to (\ref{paramreln}). This only happens when two conditions are satisfied: first, $P_{r+}$ must be respectively positive or negative; second, before squaring the square root in
 (\ref{paramreln}) one has to insure that the expression to which this square root is equal is positive. Hence, the cubic equations (\ref{cubic1}) and (\ref{cubic2}) contain ``wrong
branch'' solutions, that do not solve (\ref{paramreln}).
\begin{figure}[t]
 \centering
    \includegraphics[width=5.6cm]{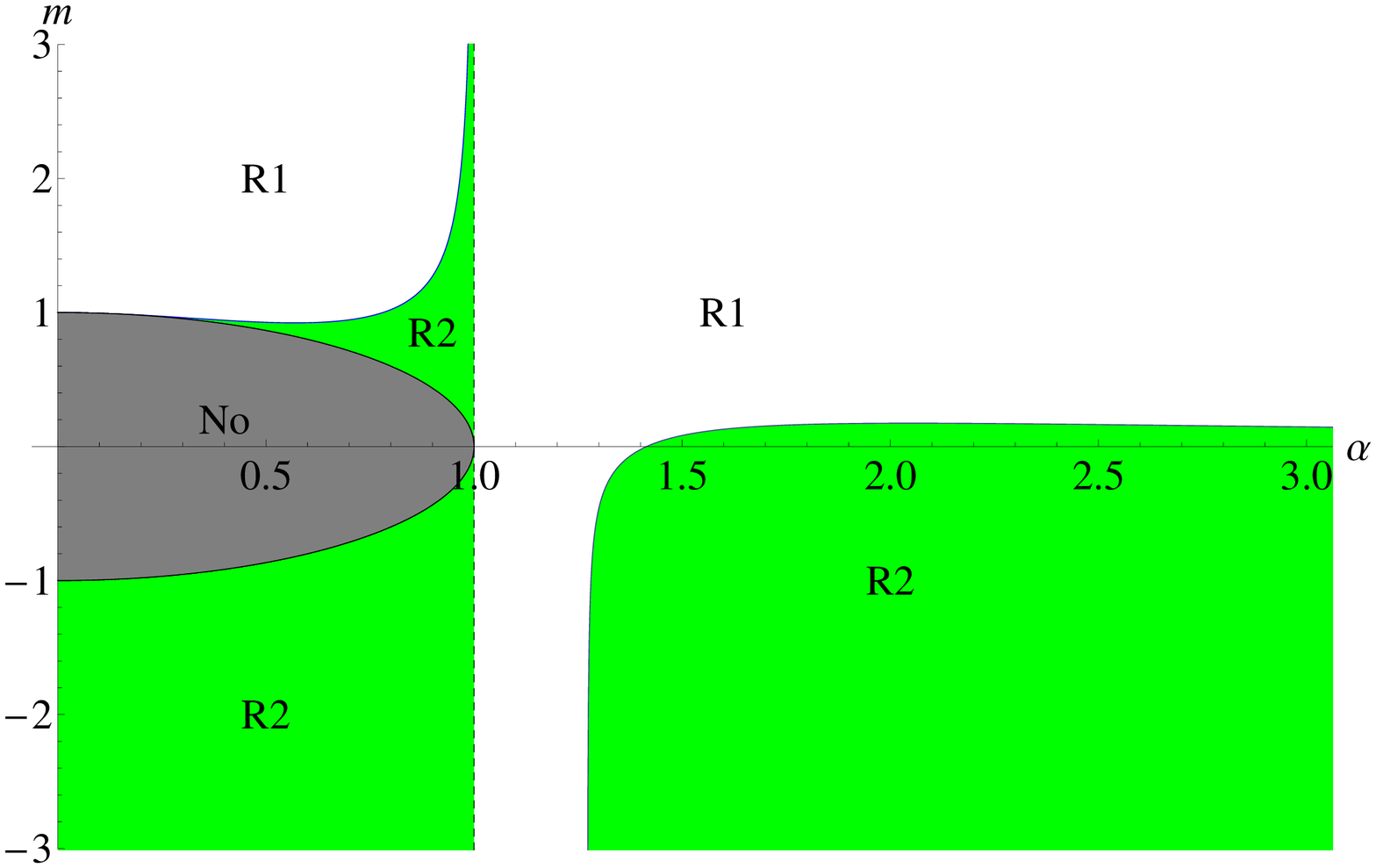}
  \hfill
    \includegraphics[width=5.6cm]{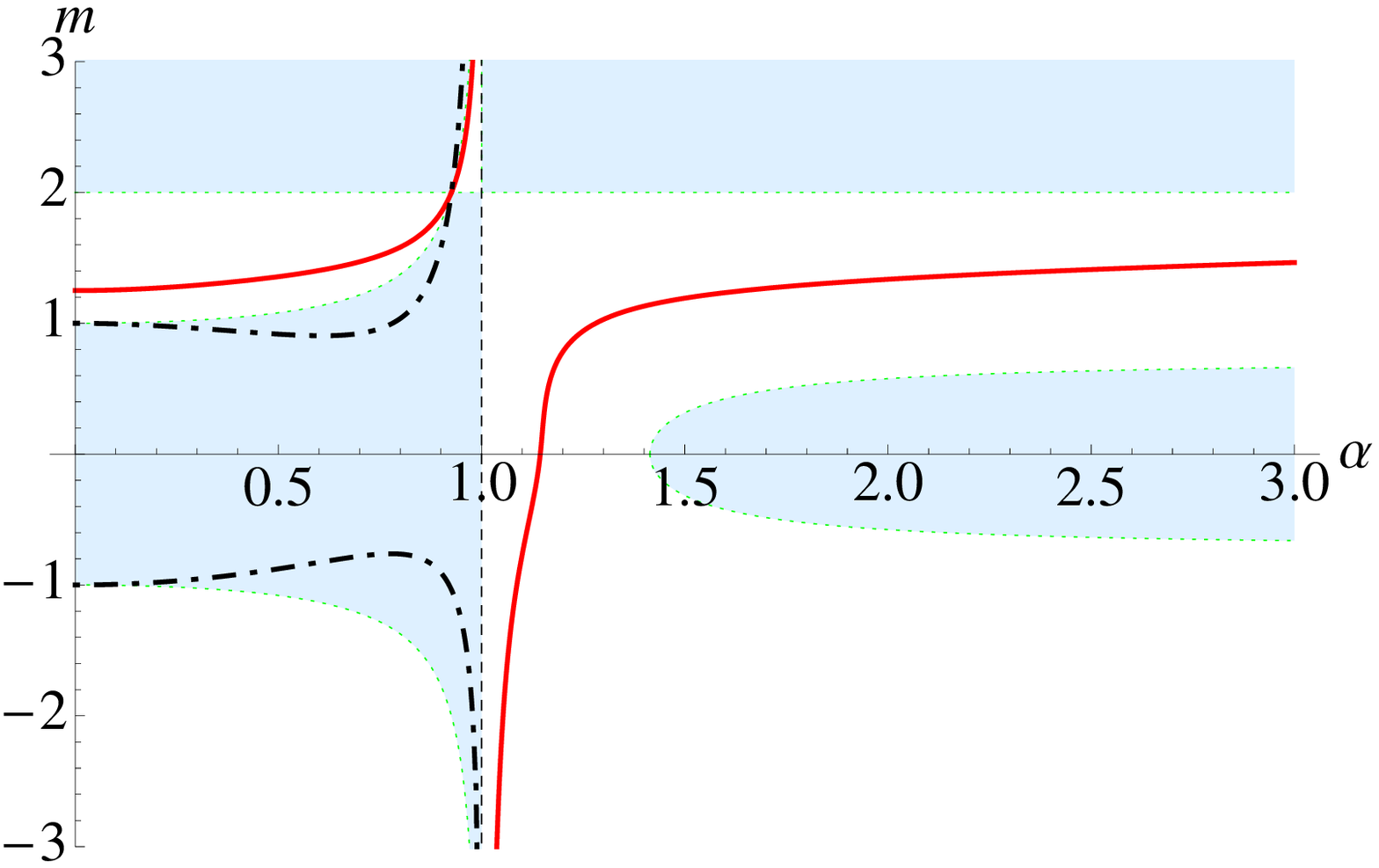}
  \hfill
    \includegraphics[width=5.6cm]{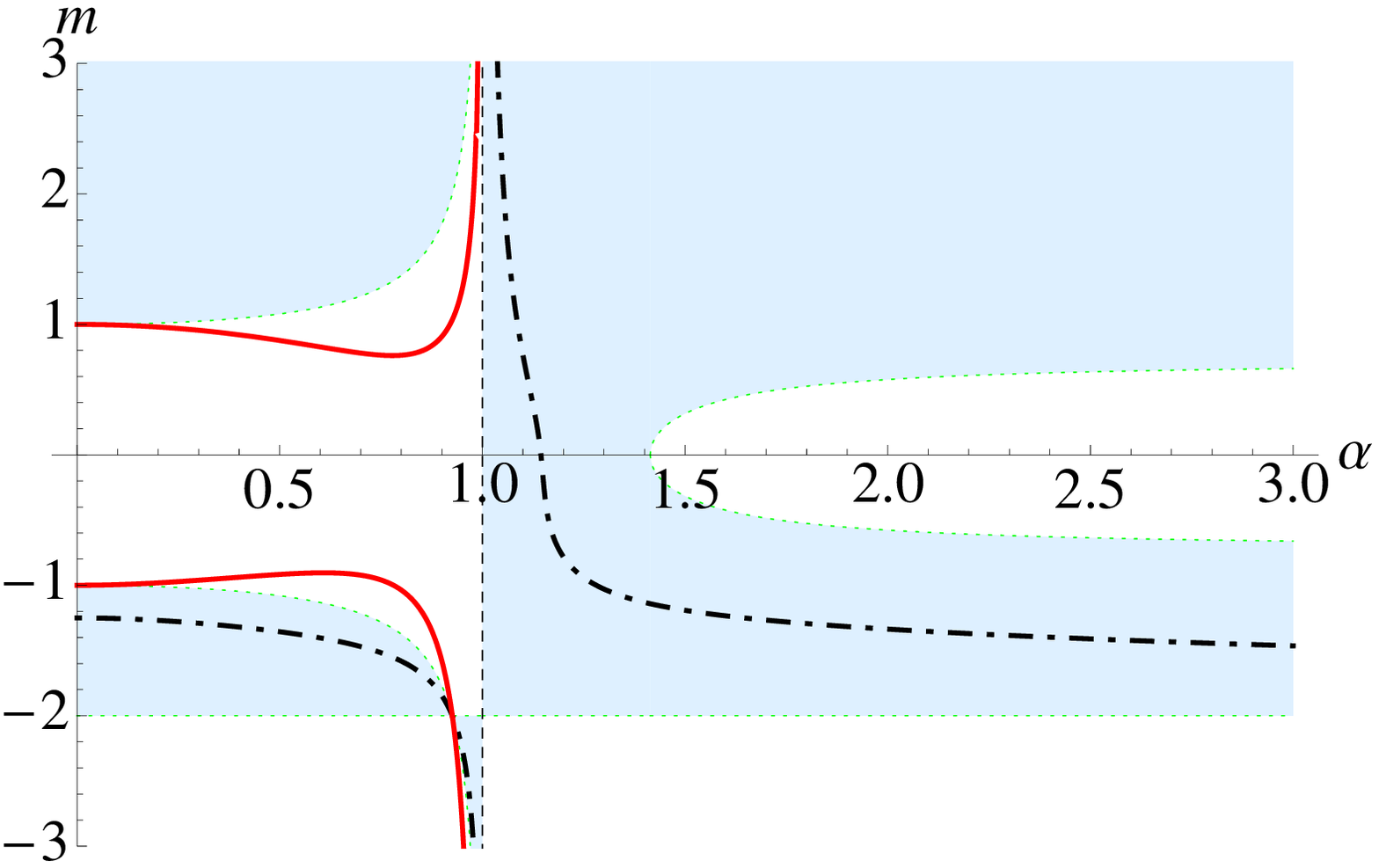}
    \caption{{\it \small The three graphs are plots of $m$ versus
        $\alpha$, in units in which $N=1$ (this choice can always be
        made because the equations are homogeneous). The first shows
        the regions where $P_{r+}$ is either positive (R1, in white)
        or negative (R2, in green).  The grey area, No, is forbidden
        by the reality condition, (\ref{rplusreal}).  The second and
        the third graph show the solutions of (\ref{cubic1}) and
        (\ref{cubic2}). In the shaded (blue) areas the square root in
        equation (\ref{paramreln}) is equal to a negative expression.
        Hence, the solutions that belong to these regions, or to the
        regions of the first graph where $P_{r+}$ has the wrong sign,
        do not obey (\ref{paramreln}) and are ``wrong branch''
        solutions. These solutions are represented using dotted lines,
        while the physical solutions, that obey (\ref{paramreln}), are
        represented using continuous lines and belong to the white
        areas.}}
\label{fig2}
\end{figure}
%
%
%
\begin{figure}[t]
 \centering
    \includegraphics[width=10cm]{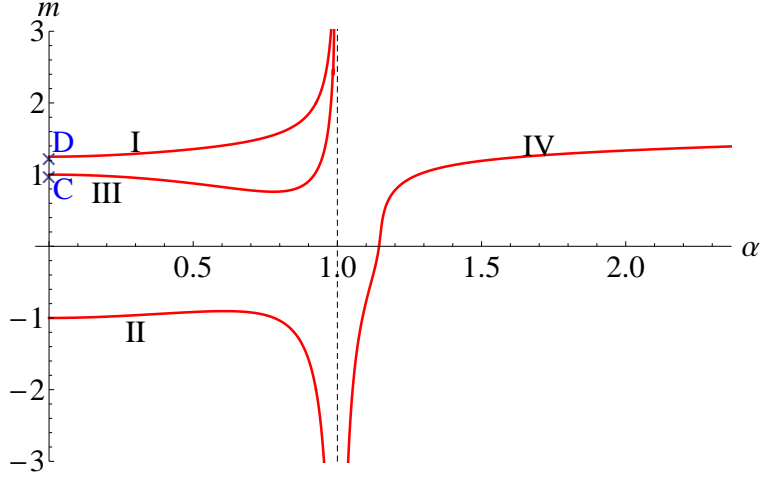}
%
    \caption{{\it \small Plot of the values of $m$ that give physical
        solutions of (\ref{paramreln}) for a given $\alpha$, in units
        in which $N=1$.  The solution has four disconnected branches:
        Branches I ,II and III go from $\alpha=0$ to $\alpha=1$,
        diverging as $\alpha$ approaches 1. Branch IV starts from
        $-\infty$ as $\alpha \rightarrow 1_+$ and approaches $m=2$ as
        $\alpha \rightarrow \infty$. The intercepts, C and D,
        correspond, respectively, to the Taub-NUT and Taub-Bolt
        metrics. }}
\label{fig3}
\end{figure}

The details of the parameter ranges are shown in Figure \ref{fig2}. The
first graph depicts the regions in which $P_{r+}$ is positive or
negative, and thus where we have to solve, respectively,
(\ref{cubic1}) or (\ref{cubic2}).  For $\alpha<1$, each cubic has
three real roots; for (\ref{cubic1}) two of them are solutions to
(\ref{paramreln}), and one lies on the ``wrong branch;'' for
(\ref{cubic2}) two of them lie on the ``wrong branch.''
For $\alpha>1$, there is one real root to the cubics,
and the only physical solution is the one with $P_{r+}>0$.

The complete solution to (\ref{paramreln}) is shown on Figure
\ref{fig3}.

We would like to note that our analysis does not completely agree with
the discussion in \cite{Gibbons:1979nf}. Indeed \cite{Gibbons:1979nf}
only analyzes solutions with positive $P_{r+}$, and thus misses some
of the ambi-polar branches of the five-dimensional solution.
Furthermore, for $\alpha<1$ the solutions found and plotted in
\cite{Gibbons:1979nf} do not have quite the same shape as the ones in
Figure \ref{fig3}\footnote{We believe this discrepancy can be most
  easily explained by evolving Moore's law backwards in time.}.

At $\alpha =0$ one obtains two interesting particular solutions:
the Taub-NUT solution, for $m=|N|$, and
the Taub-Bolt solution of \cite{Page:1979aj} for $m = 5/4 N$. It is
worth noting that allowing the metric to be ambi-polar (see section
\ref{ambipolar}) extends the range of parameters significantly.
Indeed, forcing the four-dimensional metric to have a signature
($+,+,+,+$) imposes $m>|N|$, and thus would forbid the complete branch
II, and part of branches III and IV (see Figure \ref{fig3}).

\subsection{Maxwell Fields on the Kerr-Taub-Bolt}

Introduce frames:
\begin{eqnarray}
\hat e^1 & =&    \Big( {\Xi \over \Delta} \Big)^{1 \over 2} dr \,, \qquad  \hat e^2 ~=~ \Xi^{1 \over 2} d\theta \,, \nonumber \\
  \hat e^3 & =& {\sin \theta \over \Xi^{1 \over 2}}  (\alpha d \tau + P_r d \phi)   \,, \quad \hat e^4 ~=~  \Big( {\Delta  \over \Xi} \Big)^{1 \over 2}  (d \tau + P_\theta d \phi)\,,
\label{KTBframes}
\end{eqnarray}
and define the self-dual and anti-self-dual two-forms by:
\begin{equation}
\Omega_\pm~=~ {1 \over (r \mp (N + \alpha \cos \theta))^2}   \, \big[ \hat e^1\wedge \hat e^4 \pm  \hat e^2\wedge \hat e^3\big] \,.
\label{KTBforms}
\end{equation}
These forms are harmonic and have potentials satisfying $dA_\pm = \Omega_\pm$, of the form \cite{Aliev:2005pw}:
\begin{equation}
A_\pm~=~    \mp \cos \theta \, d \phi ~-~ {1 \over (r \mp (N + \alpha \cos \theta) )}\,( d \tau + P_{\theta} d \phi) \,.
\label{KTBpots}
\end{equation}
In the rest of this section we we will focus on the self-dual Maxwell fields and take:
\begin{equation}
\Theta^{(I)} ~=~ {q_I} \, \Omega_+ \,.
\label{KTBharmtwoform}
\end{equation}
The extension to anti-self-dual Maxwell fields is rather straightforward.

Solving the second equation (\ref{BPSeqnb}) yields:
\begin{equation}
Z_I ~=~  1 ~-~ \coeff{1}{2} \, C_{IJK}  \, {q_J q_K \over  (m-N) } \, {1  \over (r - (N + \alpha \cos \theta))}  \,.
\label{KTBZI}
\end{equation}
We have, once again, chosen the homogeneous solution so as to exclude all singular electric sources for $Z_I$ and to arrange that  $Z_I \to 1$ as $r \to \infty$.  Notice that the denominator of $Z_I$ is one of the factors of $\Xi$ and, if $N \ge 0$, both this denominator and $\Xi$ will change sign when $r$ is small\footnote{We will see later that the other factor in $\Xi$ never vanishes for $N > 0$.  If $N < 0$ one can produce a similar result by starting with an anti-self-dual flux.}. This suggests that the five-dimensional metric could be  regular when the base space is ambi-polar.

\subsection{The angular momentum vector}

Solving the last equation (\ref{BPSeqnc}) is a little non-trivial and we find it convenient to make the Ansatz:
\begin{equation}
k ~=~   \mu \, (d \tau + P_\theta \, d \phi) ~+~ \nu \, d \phi  \,,
\label{kansatz}
\end{equation}
and solve the system for $\mu$ and $\nu$.  We find that this system may be recast in terms of a single function, $F$, for which:
\begin{equation}
\Xi \, \mu ~-~ \alpha \, \nu ~=~ \Delta \partial_r F \,, \qquad   \nu ~=~ \sin \theta \,    \partial_\theta F  \,.
\label{Fdefn}
\end{equation}
The equation satisfied by $F$ is:
\begin{eqnarray}
\partial_r (\Delta \partial_r F) &+& {1 \over  \sin \theta} \,\partial_\theta  (\sin \theta \,\partial_ \theta F)  - {2  \over (r - (N + \alpha \cos \theta))}  \,  (\Delta \partial_r F + \alpha \sin \theta \,    \partial_\theta F) \nonumber \\
&=& ( q_1+ q_2 + q_3) {(r +  N + \alpha \cos \theta )  \over (r - (N + \alpha \cos \theta)) }~-~ { 3 \, q_1 q_2 q_3 \over (m-N)}\,    {(r +  N + \alpha \cos \theta )  \over (r - (N + \alpha \cos \theta))^2}  \,.
\label{Feqn}
\end{eqnarray}
Upon solving this equation we find the following solution for the angular momentum vector:
\begin{eqnarray}
\mu  &=&  \gamma\, \bigg[ 1 ~-~  {2\, N  \over  (r + N + \alpha \cos \theta) } \bigg]  ~-~   (q_1+ q_2 + q_3)\, {r \over \Xi} \\
&& \qquad  ~+~{ q_1 q_2 q_3 \over 2 (m-N)^2}\,   \bigg[ {m - N - 2\, \alpha \cos \theta  \over \Xi}  ~+~ {2 (m-N)  \over (r - (N + \alpha \cos \theta))^2} \bigg]  \,, \\
\nu  &=&\gamma\,\alpha \,  \sin^2 \theta  ~-~ { \alpha \, q_1 q_2 q_3 \over (m-N)^2}\,    { \sin^2 \theta  \over (r - (N + \alpha \cos \theta)) }  \,,
\label{munusol}
\end{eqnarray}
where $\gamma$ is an arbitrary constant that multiplies terms coming from the homogeneous solution of (\ref{Feqn}). As with the Schwarzschild solution, we suppress homogeneous solutions that lead to Dirac strings (and hence CTC's) in the $\phi$ direction.

The parameter, $\gamma$, is now fixed by making sure that there are no CTC's  near $r = r_+$.  Since $\Delta$ vanishes at $r= r_+$, one can make the spatial part of the metric vanish by moving on the circle with $d \phi = -\alpha P_r^{-1} d\tau$.  It follows that, to avoid CTC's, the angular momentum vector,   $k$,  must vanish on this circle at $r= r_+$  This means that we must impose $\Xi \mu - \alpha \nu = 0$ at $r=r_+$, for any value of $\theta$.   This would follow  from the first equation in (\ref{Fdefn}) provided that $F$ has no singularity at $\Delta =0$. However, $F$ generically has terms proportional to $\log \Delta$.  On the other hand, the homogeneous solution to (\ref{Feqn}) (that yields the terms proportional to $\gamma$ in (\ref{munusol})) also contains such terms:
\begin{equation}
F_{hom} ~=~  \gamma ( r - \alpha \cos \theta)  + {(m-N) \over \sqrt{m^2 -N^2 + \alpha^2}}   (r_+ \log(r-r_+)  - \,r_- \log(r-r_-) ) \,.
\label{Fhom}
\end{equation}
Hence, we can cancel the singular behavior at $r= r_+$ by choosing the coefficient of the homogeneous solution:
\begin{equation}
\gamma ~=~   {(q_1+ q_2 + q_3)   \over 2 \, (m-N)}  ~+~ { q_1 q_2 q_3 \over 4\, (m-N)^3}\,\Big[{m+N \over r_+} ~-~ 2 \Big]  \,.
\label{gammares}
\end{equation}
The full non-singular solution than has:
\begin{eqnarray}
\mu  &=&   (\Delta + \alpha^2 \sin^2 \theta) \, \bigg[ { (q_1+ q_2 + q_3) \over 2 \, (m-N) \, \Xi} ~+~ {3\, q_1 q_2 q_3 \over 2\, (m-N)^3 \Xi } \, \Big({  N \over r_+ }  - {m-N \over 2\, r}  -1 \Big) \nonumber \\
& & \qquad \qquad  \qquad   ~-~ {q_1 q_2 q_3 \over 2\, (m-N)^2  } \,  {1 \over r\, (r - (N + \alpha \cos \theta))^2}  \bigg] \nonumber \\
&& \qquad ~-~ {3\, q_1 q_2 q_3 \over 4\, (m-N)^2  } \, \Big(1 ~+~  {2 \, N \over  (r - (N + \alpha \cos \theta)) } \Big)\,\Big( {1 \over r} ~-~{1 \over r_+} \Big) \\
\nu  &=& \alpha \,  \bigg[ {(q_1+ q_2 + q_3)   \over 2 \, (m-N)}  ~-~  { q_1 q_2 q_3 \over (m-N)}\,   \bigg( { 1 \over (r - (N + \alpha \cos \theta)) }  \nonumber\\
&& \qquad  \qquad  \qquad  \qquad ~+~ { 1 \over 4\, (m-N)^2}\,\Big({m+N \over r_+} ~-~ 2 \Big)   \bigg) \bigg]\, \sin^2 \theta  \,.
\label{munures}
\end{eqnarray}
Note that at  $r=r_+$, this solution is proportional to $\sin^2 \theta$.

\subsection{Ambi-polar Kerr-Taub-Bolt} \label{ambipolar}

The Kerr-Taub-Bolt metric (\ref{KTBmet}) can be recast in the form
\be \label{KTBmet2}
ds^2_4 = V^{-1} (d\tau + P'_\theta d\phi)^2 +V \Bigl({\Delta_\theta\over \Delta} dr^2 + \Delta_\theta d\theta^2 + \Delta \sin^2\theta d\phi^2\Bigr)\,,
\ee
with
\be
\Delta_\theta= \Delta+\alpha^2 \sin^2\theta\,,\quad V= {\Xi\over \Delta_\theta}\,,\quad P'_\theta = P_\theta +\alpha {\Xi\over \Delta_\theta}\sin^2\theta\,.
\ee
In this form, we see that if $\Xi$ becomes negative, the signature changes from ($+$,$+$,$+$,$+$) to ($-$,$-$,$-$,$-$). If one is interested in four-dimensional Euclidean metrics, one must require $\Xi>0$ and thus impose $m>|N|$. However, we are interested in regular solutions of five-dimensional supergravity, and as it is well-established, such solutions can be obtained from four-dimensional base spaces that have such signature changes \cite{Giusto:2004kj,Bena:2005va,Berglund:2005vb,Saxena:2005uk}.  We now investigate this possibility in more detail, and for simplicity we will assume $N > 0$.

The five-dimensional  metric is
\be \label{5dmet}
 ds^2 = -Z^{-2}(dt+k)^2 + Z \, V^{-1} (d\tau + P'_\theta d\phi)^2 + Z\,V ds_3^2\,,
\ee
with $ds_3^2={\Delta_\theta\over \Delta} dr^2 + \Delta_\theta d\theta^2 + \Delta \sin^2\theta d\phi^2$. The only factor in $V$ that can change sign is $\Xi$:
\be
 \Xi ~\equiv~ r^2 -(N + \alpha \cos\theta)^2 = (r -(N + \alpha \cos\theta))(r +(N + \alpha \cos\theta))\,.
\ee
It is easy to see that because $N >0$, the inequality
(\ref{rplusreal}) implies that the second factor, $(r +(N + \alpha
\cos\theta))$, is always positive\footnote{It might appear that when
  $\alpha>N$ this term can also become negative. However, the range of
  $r$ is $r \ge r_+$, and one can check straightforwardly using the regularity constraints
 that this implies $(r +(N + \alpha \cos\theta))>0$.} and so $\Xi$ changes sign
when $(r -(N + \alpha \cos\theta))$ changes sign. We therefore define
\be
 \eta  ~\equiv~  (r -(N + \alpha \cos\theta))\,.
\ee
As $\eta\rightarrow 0$, we have
\bea
 \Xi &=& 2(N+\alpha \cos\theta)\,\eta \,(1 + {1 \over 2(N + \alpha\cos\theta}\,\eta) + O(\eta^3)\,, \\ \nonumber
 \Delta_\theta &=& 2(N-m)(N+\alpha\cos\theta)(1 + { N+\alpha\cos\theta-m \over (N-m)(N+\alpha\cos\theta)}\,\eta) + O(\eta^2)\,, \\ \nonumber
 Z_I &=& {C_{IJK} \over 2}{q^Jq^k \over N-m}\,{1 \over \eta} \,(1 - {C_{IJK} \over 2} {N-m \over q^Jq^K}\,\eta) +O(\eta)\,, \\ \nonumber
 \mu &=& -{q^1q^2q^3 \over N-m}\,{1 \over \eta^2} \,\left(1+\left(-{m-N-2\alpha\cos\theta \over 4(N-m)(N+\alpha\cos\theta)} + {(q^1+q^2+q^3)(N-m) \over 2q^1q^2q^3}\right)\,\eta\right) + O(1)\,.
\eea
The first possible divergences can come from the coefficient in front of the three-dimensional metric, $ZV$. But as $\eta \rightarrow 0$,
\be
 ZV = {(q^1q^2q^3)^{2/3} \over (N-m)^2} +O(\eta)
\ee
which is perfectly regular. The factor of $Z/V \sim \eta^{-2}$ in front of the fiber metric is potentially more troublesome:
\bea
 {Z\over V} &=& {(q^1q^2q^3)^{2/3} \over \eta^2} \\ \nonumber &&+ \left({(N-m)(q^1+q^2+q^3) \over 3(q^1q^2q^3)^{1/3}} - {(q^1q^2q^3)^{2/3} \over 2(N+\alpha\cos\theta)} \Big(1 + {2(N-m+\alpha\cos\theta) \over (N-m)}\Big) \right){1 \over \eta} +O(1)\,,
\eea
and thus $g_{\tau\tau}$ appears to blow up at $\eta=0$.
However, there is a similar set of terms coming from $-Z^{-2}(dt+k)^2$
and we find that these cancel both the leading and the subleading
divergences, and $g_{\tau\tau}$ has a finite value as $\eta=0$.
Finally, one can also verify that the off-diagonal terms $g_{t\tau}$
and $g_{t\phi}$ are finite at $\eta=0$.

This cancelation of singular terms and the ultimate regularity of the
metric exactly parallels the story for the bubbled BPS solutions
\cite{Giusto:2004kj,Bena:2005va,Berglund:2005vb,Saxena:2005uk}.  Thus, when $\Xi$ changes sign, the ambi-polar
base metric leads to a regular five-dimensional metric and therefore, as described earlier,
one can allow a wider range of parameters than merely $m>|N|$ and still get a regular, Lorentzian metric in five dimensions.

\subsection{The Running Taub-Bolt Solution}

If one sets the angular momentum parameter, $\alpha$, to zero then (\ref{cubic1})  requires $m = 5 N/4$ and the base becomes the Taub-Bolt space. The full solution then simplifies dramatically and has similar general features to the Schwarzschild solution described in Section 3.
Indeed, one finds that $\nu =0$ and
\begin{equation}
\mu  ~=~     { 2\, (q_1 + q_2 + q_3)\, (r-2\,N) (r-{N \over 2}) \over N\, (r^2- N^2) }   ~+~  { 2\,  q_1 q_2 q_3 \over N^3}    \,   { (r-2\,N) \, (7r^2 - 2\,Nr + 4\,N^2)  \over(r- N)^2(r+N) }    \,.  \label{TBmures} \\
\end{equation}
We can, once again, easily check the conditions that there are no CTC's.  The only danger comes from the $\tau$-circles, whose size is given by $\cM d \tau^2$, where $\cM   \equiv  Z^{-2}  \Xi^{-2}   \cQ $ and
\begin{equation}
\cQ ~\equiv~  \Xi \, \big( \Delta Z_1 Z_2 Z_3   ~-~ \Xi\, \mu^2 \big) \,.  \label{CTCtestTB} \\
\end{equation}
This is a quartic in $r$ and must remain non-negative for $r \ge r_+ = 2N$.

To analyze this in  more detail we consider equal fluxes  $q_I = q > 0$, $I=1,2,3$.   The positivity of the $Z_I$ for $r > 2N$ implies $ q \le {1 \over \, 2}N$; furthermore, the positivity of the quartic (\ref{CTCtestTB}) imposes even stronger conditions. At infinity one has
\begin{equation}
\cM ~\sim~ r^{-4} \cQ ~ \sim ~     \Big( 1  -   6 \, \Big({q \over N }  \Big)+ 14 \, \Big({q \over N }\Big)^3\Big) \Big( 1  +  6 \, \Big({q \over N }  \Big)- 14 \, \Big({q \over N } \Big)^3\Big) \,.  \label{CTCinfTB} \\
\end{equation}
For this to be positive (with $ 0 <  q <   {1 \over 2}\, N $), the two cubics in $q$ must be positive and this implies:
\begin{equation}
0 < {q\over N}  ~< ~ {2 \over \sqrt{7}} \, \cos \Big( { \pi \over 3} + { 1 \over 3} \, \arccos\Big( {\sqrt{7} \over 4}  \Big)\Big) ~\approx~ 0.180355\,.  \label{chcondTB} \\
\end{equation}
One can then verify that (\ref{CTCinfTB}) is indeed positive definite for $2N < r < \infty$ for $q$ in the range (\ref{chcondTB}).

Since the angular momentum parameter can be viewed as a  perturbation of the foregoing solution, we anticipate qualitatively-similar solutions at least for when $\alpha$ is small.

\subsection{Asymptotic charges}

The computation of asymptotic charges proceeds along lines analogous to those outlined in Section
\ref{sec:charge}.

The solution has M5 charges encoded in the self-dual Maxwell fields $\Theta^{(I)}$ and equal to $q_I$.
The M2 charges, defined as in (\ref{QI}), are
\be
Q^I = -(8\pi N)(4\pi)\coeff{1}{2}C_{IJK}\Bigl[{q_J q_K\over m-N} + {\gamma\over 2} (q_J+q_K)\Bigr]\,,
\ee
with the parameter $\gamma$ given  in (\ref{gammares}).

As for the running Schwarzschild bolt in section 3, $\mu$ goes to a finite non-zero value, $\gamma$, at infinity. To find the mass of the solution one must then introduce  coordinates $\hat\tau$ and $\hat{t}$ as in (\ref{hatcoord}).
It is also convenient to use the form (\ref{KTBmet2}) for the Kerr-Taub-Bolt metric and to rewrite the
one-form $k$ as
\be
k=\mu (d\tau + P'_\theta d\phi) +\nu' d\phi\,,
\ee
with
\be
\nu'=\nu-\alpha  {\Xi\over \Delta_\theta}\sin^2\theta \mu =
 \alpha {q_1 q_2 q_3 (m+N)\over 2 (m-N)^2 \Delta_\theta}\Bigl(1-{r\over r_+}\Bigr) \sin^2\theta\,.
\ee
One can then rewrite the five-dimensional metric in a form ready for Kaluza-Klein reduction along $\hat\tau$:
\be
ds^2={\hat{I}_4\over (Z V)^2}\Bigl(d\hat\tau + \hat{P}_\theta d\phi - {\mu V^2\over \hat{I}_4} (d\hat{t}+\hat{\nu} d\phi)\Bigr)^2+{V Z\over \hat{I}_4^{1/2}} ds^2_E\,,
\ee
where
\be
ds^2_E = -\hat{I}_4^{-1/2}(d\hat{t}+\hat{\nu} d\phi)^2 + \hat{I}_4^{1/2} \Bigl({\Delta_\theta\over \Delta} dr^2 + \Delta_\theta d\theta^2 + \Delta \sin^2\theta d\phi^2\Bigr)
\ee
is the four-dimensional Einstein metric and
\bea
\hat{I}_4 =  (1-\gamma^2)^{-1}(Z_1 Z_2 Z_3 V - \mu^2 V^2)\,,\quad \hat{P}_\theta = (1-\gamma^2)^{1/2} P'_\theta\,,\quad \hat{\nu}=(1-\gamma^2)^{-1/2} \nu'\,.
\eea
From this one can read off the mass, $M$ and the four-dimensional angular momentum, $J$:
\bea
G_4 M &=& {1\over 4 (1-\gamma^2)} \Bigl[2m - {q_1 q_2 + q_1 q_3 + q_2 q_3\over m-N}\nonumber\\
&&\qquad+{q_1 q_2 q_3 (q_1 +q_2 +q_3)\over 2 (m-N)^3}\Bigl(2-{m+N\over r_+}\Bigr)-{(q_1 q_2 q_3)^2 \over 4 (m-N)^5} \Bigl(2-{m+N\over r_+}\Bigr)^2\Bigr]\,,\\
G_4 J &=& -{1\over (1-\gamma^2)^{1/2}}{\alpha q_1 q_2 q_3 (m+N)\over 4 (m-N)^2 r_+}\,.
\eea
The geometry also carries Kaluza-Klein electric charge, $Q_e$, and the Kaluza-Klein magnetic charge, $Q_m$, given by\footnote{We use the conventions of \cite{Elvang:2005sa}.}
\bea
G_4 Q_e&=&-{1\over 8 (1-\gamma^2) (m-N)^3}\Bigl[q_1 q_2 q_3 \Bigl({m\over 2} - 2 N + {N^2-m^2\over r_+}\Bigr) \nonumber\\&&+ {1\over 2} (q_1 q_2 +q_1 q_3 + q_2 q_3) (q_1 +q_2 + q_3) (m+2 N)+{1\over 2} (q_1^3+q_2^3+q_3^3) m \nonumber\\
&&- {1\over 2} q_1 q_2 q_3(q_1 q_2 +q_1 q_3 + q_2 q_3) {m+2N\over (m-N)^2} \Bigl(2-{m+N\over r_+}\Bigr)\nonumber\\
&& -{1\over 4} q_1 q_2 q_3 (q_1^2+q_2^2+q_3^2){2m+N\over (m-N)^2} \Bigl(2-{m+N\over r_+}\Bigr) \nonumber\\
&&+{1\over 8} q_1^2 q_2^2 q_3^2 (q_1+q_2+q_3) {m+2N\over (m-N)^4}  \Bigl(2-{m+N\over r_+}\Bigr) ^2\nonumber\\
&&-{1\over 16} q_1^3 q_2^3 q_3^3 {N\over (m-N)^6}  \Bigl(2-{m+N\over r_+}\Bigr)^3  \Bigr]\,,\\
G_4 Q_m &=& (1-\gamma^2)^{1/2} {N\over 2}\,,
\eea
where $G_4$ is the four-dimensional Newton's constant
\be
G_4 = {G_5\over (1-\gamma^2)^{1/2} (8\pi N)}\,.
\ee

\section{Conclusions}

We have constructed a five-parameter family of smooth horizonless
solutions of five-dimensional $U(1)^3$ ungauged supergravity (or of
the STU model) that are asymptotically $\IR^{3,1} \times S^1$ and
that can have the same charges as a five-dimensional black string with
M5, M2 and KK momentum charges and a macroscopically-large horizon
area.  We find that in these solutions the KK fiber and the time
always mix at infinity and, since they contain a bolt, we refer to
them as ``running Bolt'' solutions.

As these solutions exist in the same regime of parameters as the
classical black hole, perhaps the most important question raised by
their existence is whether they should be thought of as microstates of
this black hole, and consequently whether their physics can be used to
support the extension of the fuzzball proposal (also known as Mathur's
conjecture) to non-extremal black holes, and in particular to the
Schwarzschild black hole. As explained in the Introduction, this would
be the first example of string theory resolving a spacelike
singularity -- and the resolution mechanism would be rather amazing,
as it would imply that the resolved spacetime differs from the
singular spacetime on a macroscopic scale that is much larger than the
Planck or string scale.

For black holes that have an $AdS$ throat it is, more or less,
straightforward to argue that a certain smooth geometry is a black
hole microstate. By the $AdS$-CFT correspondence such a geometry must be
dual (up to $1/N$ corrections) to a coherent pure state of the dual
CFT. Since the entropy of the CFT is reproduced by the entropy of the
black hole \cite{Strom-Vafa}, the bulk dual of one of the CFT
microstates can be properly thought of as one of the black hole
microstates. On the other hand, if a certain black hole does not have
an $AdS$ throat, the only way one can ``define'' its microstates is by
requiring that they be smooth, horizonless, and have the same charges,
mass and angular momentum.

The running Bolt solutions we construct here satisfy this criterion.
On the other hand, it is clear that, much like the other non-extremal
solutions constructed so far, they are quite different from the
geometries that should correspond to the typical microstate of a
non-extremal black hole. For example a point particle placed in any
``Euclidean instanton times time'' solution does not feel any
attractive force. Similarly, an M2 brane probe placed in the running
Bolt geometries does not feel any force. On the other hand, both the
non-extremal black hole solution, and any typical microstate thereof
will attract such probes.

It is clearly important to explore in more detail the relation between
our running Bolt geometries and the black hole solutions that have the
same charges. The latter are rotating four-dimensional non-extremal
black holes with D4, D2 and D0 charge, and to our knowledge their
supergravity solution is not known. When uplifted to five dimensions
these black holes will become black strings with M5, M2 and KK
momentum charges, and, at least for small-enough charges, they might
suffer from Gregory-Laflamme-type instabilities
\cite{Greg-Laf}\footnote{Although it is quite possible that M5 branes
  along the putative Gregory-Laflamme direction might prevent this
  instability to occur even for arbitrarily-small M5 charges.}. On the
other hand, when the charges become large these instabilities will
most likely disappear, as they do for charged black strings
\cite{Greg-Laf2}. It would be quite important to establish this, and if
both the black hole and the microstates are unstable to compare their
decay times.

Another important calculation is that of the decay time of our
solutions, and its (rather non-trivial) dependence on the electric and
magnetic charges and the angular momentum. Again, when these charges
are zero, the Euclidean Schwarzschild \cite{GPY} and the Euclidean
Taub-Bolt solutions \cite{Young:1983dn} have a negative-mass mode, and
hence the five-dimensional solution coming from adding a time
direction to these solutions is unstable. It is unlikely that a small
amount of flux on the bolt will change this, but as the D2 and D4
charges grow the running Bolt solution may or may not become stable.
If one knew the dependence of the decay time on charges, and if one
used the fact that the non-extremal rotating black holes with D4-D2-D0
charges have a microscopic description in terms of an MSW string
\cite{MSW} with both left- and right-movers excited \cite{Finn}, one
could try looking for a microscopic state whose decay time depends in
the same non-trivial way on the charges -- this will give us the dual
of these geometries, and indicate their degree of typicality.

An interesting aspect of the running Bolt solutions we construct is
the relation between their mass and their charges. When no charges are
present, the mass of the five-dimensional uncharged Bolt solutions
obtained by adding time to a four-dimensional Euclidean instanton is
proportional to ${1 \over g^2}$, where $g$ is the four-dimensional
coupling constant coming from the KK reduction from five to four
dimensions. As such, the uncharged Bolt solutions have a solitonic
character. When charges are added, the KK fiber and the time always
mix by a non-zero amount at infinity, and the bolt starts to run. If
one computes its rest mass, one finds that when the base space is
Euclidean Schwarzschild and the fluxes are self-dual, this mass can be
written as a sum of the mass of the uncharged Bolt and of the three M2
brane charges. Hence, although this solution is non-BPS, if one
naively ascribed a ``solitonic charge'' to the stationary Bolt, the
mass of the self-dual running Bolt could be written in a BPS fashion,
as the sum of its M2 and ``solitonic'' charges.

On the other hand, we have found that the rest mass of anti-self-dual
running Bolts is given by the solitonic mass {\it minus} the M2
charges. Hence, the mass of the anti-self-dual running Bolt decreases
linearly with the increasing of its M2 charges! We could not find
other examples of such an unusual behavior of a mass formula in the
literature; this comes essentially from the fact that the constant
and $r$-dependent part of the warp factors have opposite signs. We
have furthermore shown in Section \ref{remarks} that this rather
unusual mass formula, combined with the existence of M2 brane probes
that feel no force does not violate energy conservation, as one might
naively have expected.

One could worry that the decreasing of the energy with
the M2 charges might violate the positive energy theorem. However,
since the KK circle pinches off at the bolt, the fermions must be antiperiodic around this circle, and this is incompatible with supersymmetry. Therefore the standard positive-energy theorems do
not apply \cite{Witten:1981gj,Horowitz:2005vp}\footnote{We thank Harvey Reall for pointing this argument to us.}.
 It is clearly interesting to extend our
analysis of the rest mass to the  running Kerr-Taub-Bolt solution,
and see whether for anti-self-dual fluxes the rest mass of that
solution also decreases with increasing charge. It would be even more
interesting to find an explanation for this phenomenon.

Finally, the solutions we construct, as well as the ones of
\cite{Jejjala:2005yu,Giusto-Ross-Saxena,AlAlawi:2009qe} represent a
very tiny portion of the expected smooth horizonless microstate
geometries for non-extremal black holes. Constructing large families
of such geometries in a systematic fashion is a challenging, yet very
worthwhile goal.  The nature of these geometries is not more
complicated than that of multiple concentric black rings, and we
believe it should be possible to generalize the rod-structure and
inverse-scattering methods used in their construction \cite{inverse}
to (at least) minimal five-dimensional ungauged supergravity, use one
of the known solutions as a seed, and construct the multiple-bubble
equivalent of running Bolts or of the geometries of
\cite{Jejjala:2005yu,Giusto-Ross-Saxena,AlAlawi:2009qe}. One could
also try constructing multiple-bubble non-extremal solutions by
turning on fluxes on multi-Schwarzschild solitons.  As one finds with
the BPS solutions, the presence of fluxes might compensate the mutual
attraction of the bolts, and lead to a five-dimensional smooth
solution despite the presence of struts on the four-dimensional base
space. Needless to say, if such large families of multiple-bubble
solutions were found, and if their physics equally supported the
extension of the fuzzball proposal of non-extremal black holes, this
will greatly advance our understanding of not only these black holes,
but also of the way in which string theory resolves space-like
singularities.

\bigskip
\leftline{\bf Acknowledgments}
\smallskip
We would like to thank Gary Horowitz, Malcolm Perry and Harvey Reall for interesting
discussions and suggestions.  NPW is grateful to the IPhT(SPhT), CEA-Saclay for
hospitality while this work was done. The work of IB, CR and SG was
supported in part by the DSM CEA-Saclay, by the ANR grants BLAN
06-3-137168 and 08-JCJC-0001-01 and by the Marie Curie IRG 046430.
The work of NPW was supported in part by DOE grant DE-FG03-84ER-40168.
NPW would like to dedicate this work to Abbey, a very dear friend over
the last 12 years and with whom he had many valuable conversations
about black holes.



\end{document}